\documentclass[%
 aip,
 rsi,
 amsmath,amssymb,
 reprint,%
]{revtex4-1}

\usepackage{graphicx}
\usepackage{dcolumn}
\usepackage{bm}

\usepackage[utf8]{inputenc}
\usepackage[T1]{fontenc}
\usepackage{mathptmx}
\usepackage{hyperref}

\usepackage{xcolor}

\begin{document}

\preprint{AIP/123-QED}

\title[Real-time Thomson scattering evaluation on LHD]{Initial operation and
    data processing on a system for real-time evaluation of Thomson 
    scattering signals on the Large Helical Device}

\author{K.C.~Hammond}
  \email{khammond@pppl.gov}
\author{F.M.~Laggner}
\author{A.~Diallo}
\author{S.~Doskoczynski}
\author{C.~Freeman}
  \affiliation{Princeton Plasma Physics Laboratory, Princeton, NJ 08543, USA} 
\author{H.~Funaba}
  \affiliation{National Institute for Fusion Science, Toki 509-5292, Japan}
\author{D.A.~Gates}
\author{R.~Rozenblat}
\author{G.~Tchilinguirian}
  \affiliation{Princeton Plasma Physics Laboratory, Princeton, NJ 08543, USA} 
\author{Z.~Xing}
  \affiliation{Princeton University, Princeton, NJ 08544, USA}
\author{I.~Yamada}
\author{R.~Yasuhara}
  \affiliation{National Institute for Fusion Science, Toki 509-5292, Japan}
\author{G.~Zimmer}
  \affiliation{Princeton Plasma Physics Laboratory, Princeton, NJ 08543, USA} 
\author{E.~Kolemen}
  \affiliation{Princeton Plasma Physics Laboratory, Princeton, NJ 08543, USA} 
  \affiliation{Princeton University, Princeton, NJ 08544, USA}



\date{\today}

\begin{abstract}
A scalable system for real-time analysis of electron temperature and density
based on signals from the Thomson scattering diagnostic, initially developed 
for and installed on the NSTX-U experiment, was recently adapted
for the Large Helical Device (LHD) and operated for the first time during 
plasma discharges. During its initial operation run, it routinely recorded
and processed signals for four spatial points at the laser repetition rate
of 30 Hz, well within the system's rated capability for 60 Hz. 
We present examples of data collected from this initial run and describe 
subsequent adaptations to the analysis code to improve the fidelity of the 
temperature calculations.
\end{abstract}

\maketitle

\section{Introduction}
\label{sect:intro}

The ability to determine plasma temperature and density profiles 
in real time is valuable to plasma control systems seeking to optimize
plasma properties or maintain device safety. The Thomson scattering diagnostic
is an attractive candidate for real-time profile evaluation due
to its ability to make non-invasive measurements of electron temperature $T_e$ 
and electron density $n_e$ throughout a typical fusion 
plasma.\cite{hutchinson2002}
In addition, the localized nature of the measurement obviates the need for
intermediate analysis such as tomographic inversion or equilibrium modeling
to infer the profiles.

Systems for evaluating Thomson scattering data in 
real time have been developed for a number of devices over the 
years.\cite{greenfield1990a,shibaev2010a,shibaev2010b,kolemen2015a,
arnichand2019a} In this work, we use a framework that was developed recently
for NSTX-U\cite{laggner2019a} but has not yet been tested during 
plasma discharges. Its main components
are a set of fast digitizers and a multi-core server. The digitizers 
collect signals from the detection electronics.
The server employs real-time software to calculate $T_e$ and $n_e$ from
the signals between subsequent laser pulses, supporting repetition rates
of up to 60~Hz.\cite{rozenblat2019a}
The server outputs $T_e$, $n_e$, and their
uncertainties for each scattering volume as analog signals that may be fed 
into control systems.

The setup was subsequently replicated and installed on the Large Helical 
Device (LHD) to evaluate signals from the LHD Thomson scattering system.
This system currently employs four Nd:YAG lasers that enable pulse repetition 
rates ranging from 10 to 100~Hz.\cite{yamada2016a} 
Backscattered light is transmitted through optical fibers
to a set of 144 polychromators.\cite{narihara1997a,narihara2001a} 
Within each polychromator, the scattered light passes through bandpass 
filters to six avalanche photodiodes (APDs).\cite{yamada2010a} 
Signals from five of these APDs are used for $T_e$ and $n_e$ evaluation during 
plasma discharges, and the remaining signal is used for Rayleigh 
calibration.\cite{yamada2007a} As the setup is
similar to what is employed in NSTX-U,\cite{leblanc2003a,diallo2012a} 
the real-time framework could be adapted for LHD with minor 
modifications.

During its initial operation on LHD, the real-time system 
collected and evaluated signals from four polychromators. For these
experiments, laser pulses occurred at a 30~Hz repetition rate. 
In addition to performing 
$T_e$ and $n_e$ calculations for each pulse, the real-time system archived the
polychromator signals at the end of each discharge, thereby enabling
a postmortem assessment of the real-time calculations. In this paper, we
describe this assessment, with a focus on the methods used for processing
the raw polychromator signals. In Sec.~\ref{sec:sigEval}, we assess three
signal-processing methods.
Then, in Sec.~\ref{sec:ref_compare}, we compare the values of $T_e$ and $n_e$
derived from each of these methods with the values calculated by the standard
LHD methods performed after each discharge.

\section{Signal processing methods}
\label{sec:sigEval}

The real-time system computes $T_e$ and $n_e$ values in a least-squares
spectral fitting technique
that optimizes $T_e$ and a scaling factor $c$ to minimize $\chi^2$, 
defined as follows:

\begin{equation}
    \chi^2 = \sum_{i=1}^{N_c} \frac{\left(s_i - cF_i(T_e)\right)^2}
                                   {\sigma_{s_i}^2 + s_i^2\sigma_{F_i}^2}
    \label{eqn:chi2}
\end{equation}

\noindent Here, $N_c$ is the number of polychromator channels (five for LHD)
and $F_i(T_e)$ is the expected relative signal strength from the $i$th channel
for electron temperature $T_e$. The electron density $n_e$ can then be 
determined from $c$ along with calibration data. $s_i$ is a measure of the
signal strength from the $i$th polychromator channel. $\sigma_{s_i}$ is the
standard error in the $s_i$ measurement, and $\sigma_{F_i}$ is the 
(dimensionless) fractional error in $F_i$.

\begin{figure}
    \begin{center}
    \includegraphics[width=0.48\textwidth]{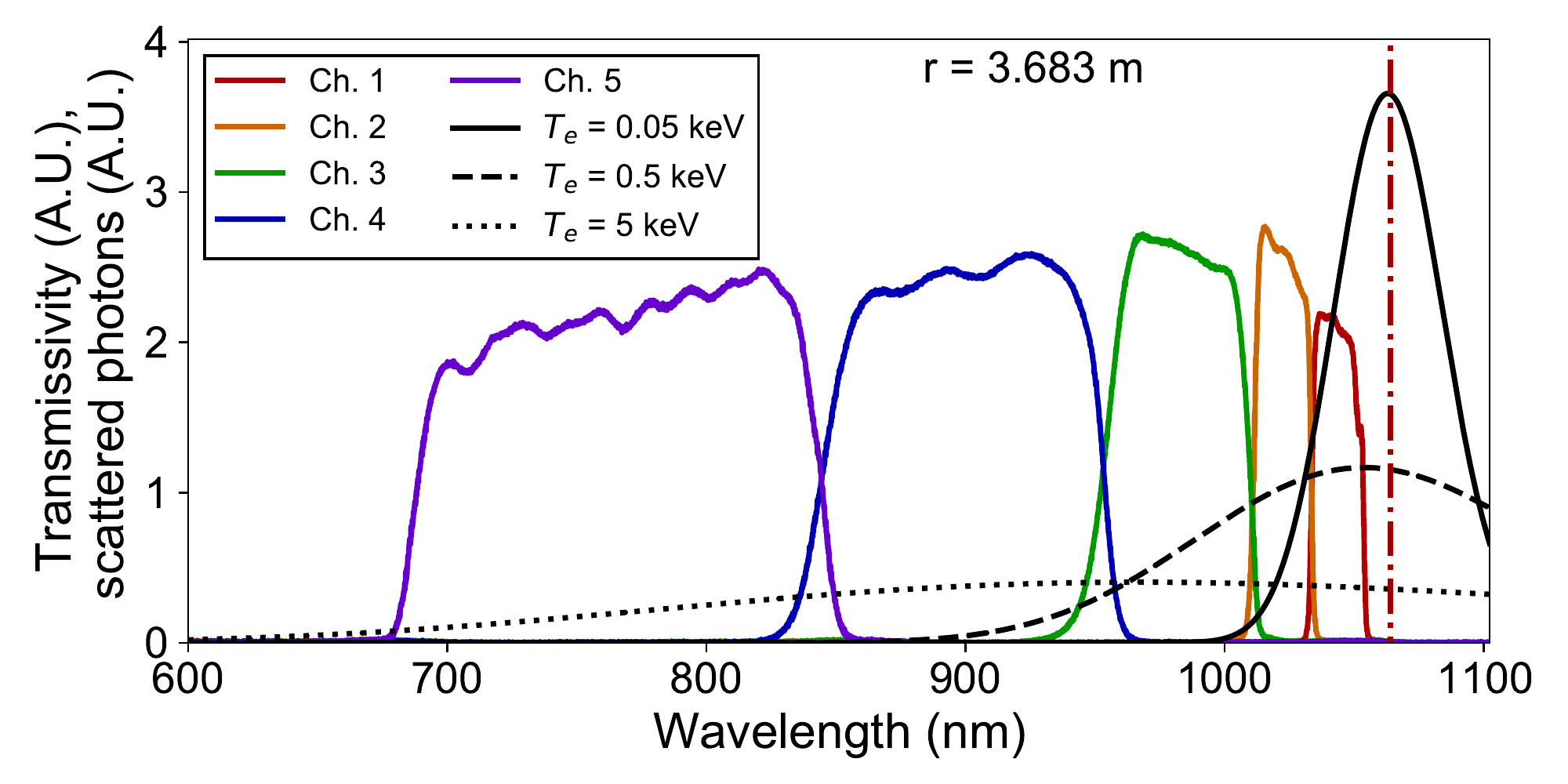}
    \caption{Transmissivity spectra of the five channels of the polychromator
             surveying the scattering volume at $r=3.683$~m, along with
             Thomson scattering spectra for three example electron 
             temperatures.}
    \label{fig:spec_calib}
    \end{center}
\end{figure}

As described in 
Ref.~\onlinecite{laggner2019a}, the values of $F_i$ are pre-computed for a 
range of $T_e$ by integrating the measured transmissivity spectrum of 
the $i$th polychromator channel with the Thomson scattering spectrum as
predicted by the Selden model.\cite{selden1980a} Fig.~\ref{fig:spec_calib}
shows transmissivity curves for one polychromator,
along with scattering spectra for selected values of $T_e$. Evaluation 
of $F_i(T_e)$ between laser pulses entails querying a table of
the pre-computed values. 

\begin{figure*}
    \begin{center}
    \includegraphics[width=\textwidth]{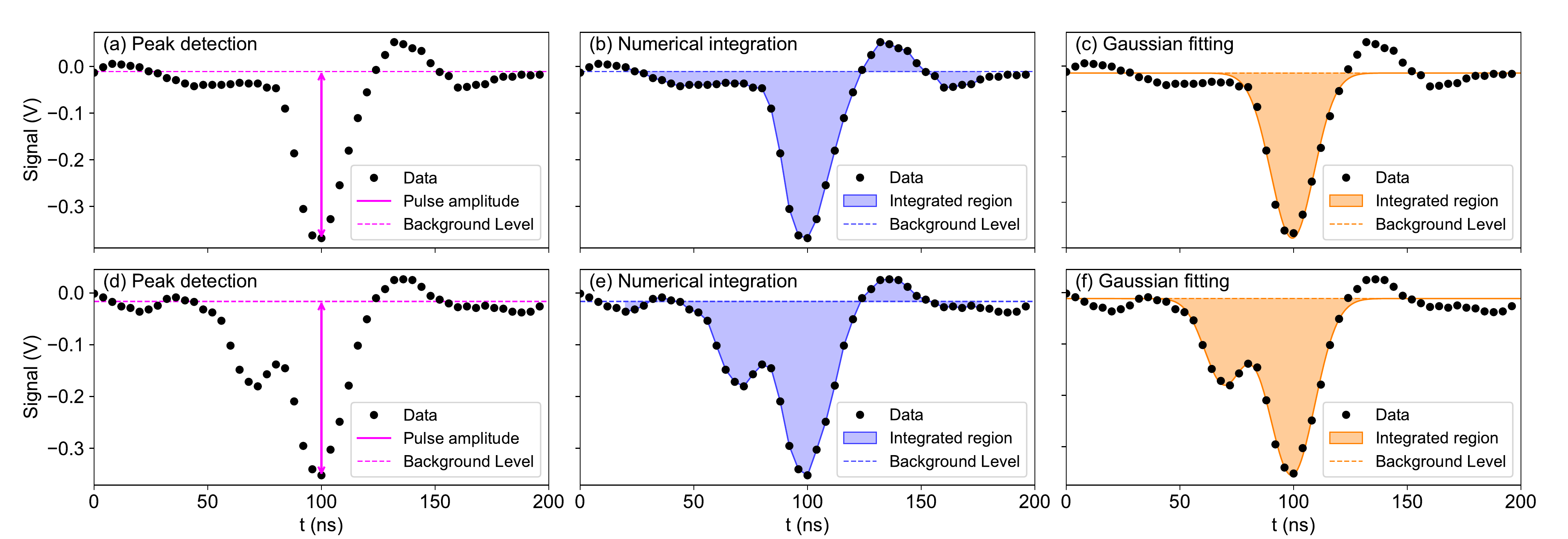}
    \caption{Typical signal from a polychromator channel along with depictions 
             of the three methods for evaluation considered in this work: 
             (a) peak detection, 
             (b) numerical integration, and
             (c) Gaussian curve-fitting for a signal arising from light 
                 scattered from a single laser pulse;
             (d)-(f) like (a)-(c) but for a signal arising from light scattered
                 from two overlapping laser pulses.}
    \label{fig:pulseEval}
    \end{center}
\end{figure*}

$s_i$ in Eq.~\ref{eqn:chi2} must be proportional to the number of
scattered photons received by the i$^{th}$ channel. 
In this section, we will describe three procedures 
for deriving $s_i$ from polychromator 
signals and compare their 
accuracy and computational requirements. The procedures, which include
peak detection, direct integration, and curve fitting, are displayed 
schematically in Fig.~\ref{fig:pulseEval} with examples of two types of
raw signals. The first type 
(Fig.~\ref{fig:pulseEval}a-c) is a response to the firing of a 
single laser and will be categorized as a ``single laser pulse,'' whereas
the second type (Fig.~\ref{fig:pulseEval}d-f) arises from
two lasers firing in rapid succession along the same beam path and will be
categorized as a ``double laser pulse.''

Each procedure was implemented as a C++11 function that could be called by
the real-time software. The functions each took a digitized raw signal as 
input and returned values for $s_i$ and $\sigma_{s_i}$. The computational
expense of each procedure was determined by recording the time required
for its respective function to process archived experimental signals.
These tests were performed on the same type of processing
hardware employed in the experimental setup, i.e. a SuperMicro real-time 
server with 32 GB RAM and 20 cores clocked at 
2.1 GHz.\cite{laggner2019a,rozenblat2019a}

We will use function evaluation times to determine how much each 
procedure will add to the system's end-to-end processing time.
This duration, which includes digitization, signal
evaluation, plasma parameter calculation, and output signal generation, was 
previously determined in offline tests to be less 
than 16.7~ms with the peak detection procedure. While the calculations for 
each scattering volume are performed in parallel threads,
the values of $s_i$ and $\sigma_{s_i}$ for each polychromator channel are
determined in series. Hence, we will estimate the additional 
end-to-end-time required for a given procedure as the time by 
which the evaluation of the five signals from a polychromator with the
procedure exceeds that of peak detection. To accommodate
the 30~Hz laser repetition rate used during the LHD experiments, the total
time should not exceed 33.3~ms.

\subsection{Peak detection}
\label{sec:peakDetect}

As originally designed, the software worked on the assumption
that the photon flux to each polychromator channel would scale
directly with the peak amplitude of its signal.
Hence, it would suffice to define $s_i$
simply as the peak amplitude in order to determine $T_e$.
Provided that the signals from absolute calibrations are
evaluated in the same way, a consistent scaling factor may be determined
for extracting $n_e$ from $c$.

The real-time code determines the 
amplitude of the scattered-light signal during each laser pulse by
subtracting the peak value of the signal from the average of the
background signal recorded prior to the laser pulse. The procedure is
illustrated in Fig.~\ref{fig:pulseEval}a for the case of a single laser pulse 
and in Fig.~\ref{fig:pulseEval}d for a double pulse.
The background level, illustrated by the horizontal dashed line, is obtained
as the average of a group of sample points separated from the peak by at
least 80~ns. 
To correct for stray laser light due to reflections within the plasma vessel, 
this amplitude would be adjusted
by subtracting the average signal amplitude recorded during laser pulses 
prior to the beginning of the discharge from the corresponding 
scattering volume and polychromator channel.

This approach has the advantage of simplicity and speed. 
However, the amplification circuitry, which in this setup includes a high-pass
filter, may exhibit a dispersive or nonlinear
response that would give rise to distortions of the signal 
that can vary from one channel to another. Evidence of this response can
be seen in the overshoot of the signal above the background level following
the peak in the examples shown in Fig.~\ref{fig:pulseEval}.

\begin{figure}
    \begin{center}
    \includegraphics[width=0.47\textwidth]{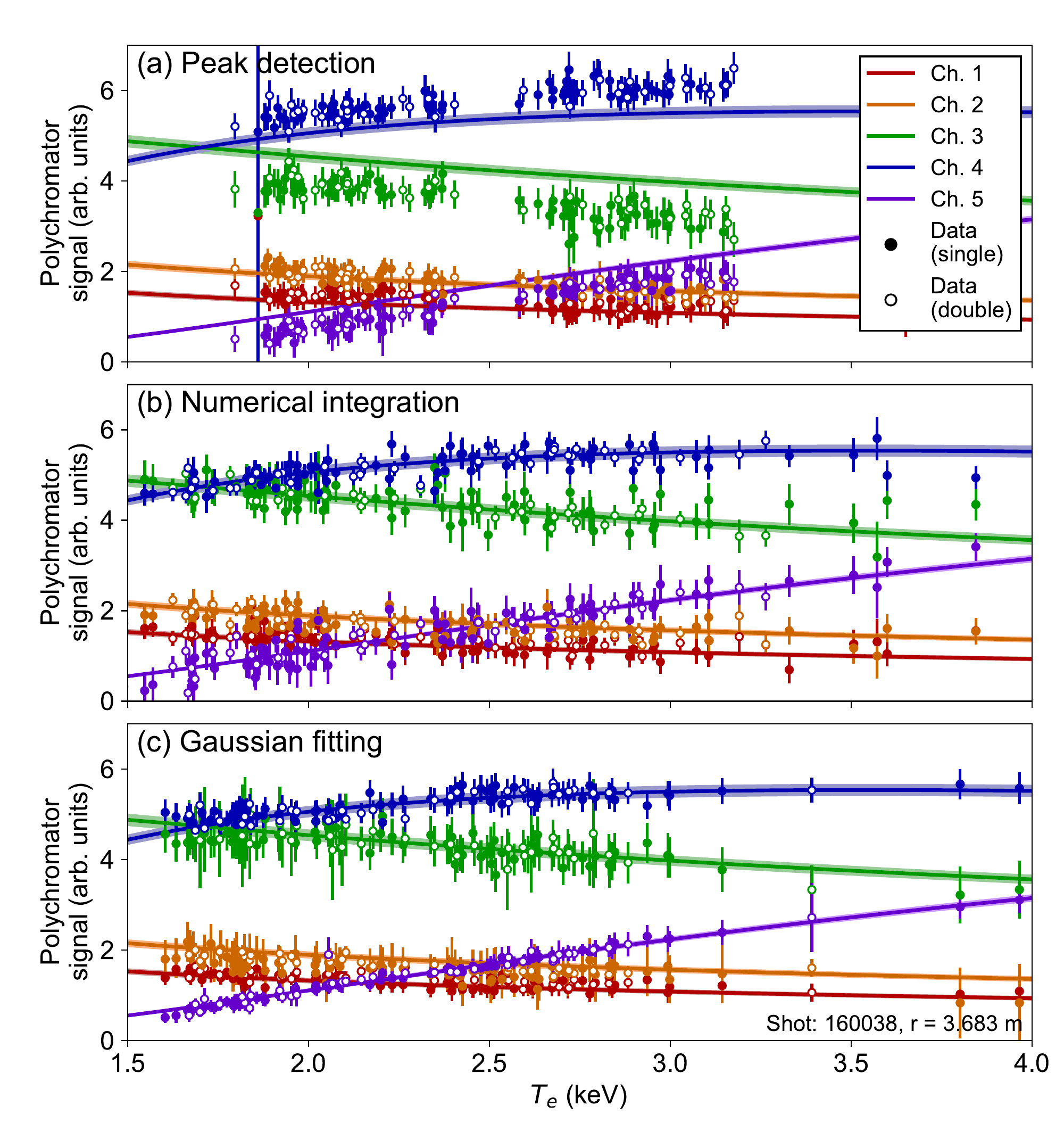}
    \caption{Comparison of calibration spectra with fitted polychromator signals
             from the scattering volume at $r=3.683$~m.
             Solid curves represent $F_i(T_e)$
             from each polychromator channel. Circles represent $s_i/c$, 
             positioned to $T_e$ according their respective fits.
             Signals determined through
             (a) peak detection;
             (b) numerical integration; 
             (c) integration of a Gaussian fit curve.}
    \label{fig:spec}
    \end{center}
\end{figure}

The effect of such distortions in the fitting of the polychromator output
spectra is visible in Fig.~\ref{fig:spec}a, which
compares fitted data $s_i/c$ with the predicted signals $F_i(T_e)$ 
acquired from one scattering volume during an example plasma discharge.
These fits contain substantial systematic errors.
In particular, the fitted signal amplitudes are mostly lower than the model
prediction for polychromator channels 3 and 5, whereas they are mostly higher 
than the model prediction for channel 4. 

\subsection{Numerical integration}
\label{sec:numInt}

In the presence of distorted polychromator signals,
the integrals of the signals
may provide a more accurate representation of the detected scattered light
than the amplitudes alone. To test this, we implemented a modified version
of the signal processing function that performs trapezoidal integration on
an interval of the signal of pre-determined length centered around the peak
value. This procedure is illustrated schematically in Fig.~\ref{fig:pulseEval}b
and \ref{fig:pulseEval}e. Analogously to the case of pulse amplitude 
determination, the integrals of signals recorded during the plasma discharge
were corrected by subtracting the average integral value of a set of stray 
laser light signals obtained prior to the plasma discharge. 

The switch to numerical integration has
minimal impact on computation time, with each integration taking
$\leq~1~\mu$s longer than a corresponding amplitude determination. Hence, to
evaluate the five polychromator outputs necessary for a spectral fit, the
added time would not exceed 5~$\mu$s. This is negligible compared to the
laser firing period of $16.7$~ms for which the system was designed.

As shown in Fig.~\ref{fig:spec}b, the use of integrated signals improved the
quality of the spectral fits. In particular, the systematic offsets between
the data and the fitted spectra that arose from the use of pulse amplitudes
(Fig.~\ref{fig:spec}a) are no longer visible. 

While the use of numerical integration removed the systematic
offsets, the fits did exhibit increased
scatter, particularly for the weaker signals from 
channel 5. This may arise from background noise, which can make spurious 
contributions to the integral in the case of a low signal-to-noise ratio (SNR)
as seen in Fig.~\ref{fig:pulseEval}b and 
\ref{fig:pulseEval}e. As the background fluctuations in the raw signal
exhibit time scales similar to the laser pulse width, their contributions 
to the integral will not necessarily cancel out.

\subsection{Curve fitting}
\label{sec:curveFit}

One way to avoid contributions from background fluctuations is to integrate
a waveform fitted to the signal rather than the 
raw signal itself.\cite{kurzan2004a,bozhenkov2017a} To this end, we implemented 
an additional function that fits Gaussian curves to the signal and outputs 
the integrals computed
analytically from the fitting parameters. Specifically, signals 
were fit to functions of the form

\begin{equation}
    f(t) = p_1 + p_2 \exp\left[-\frac{(t-p_4)^2}{2p_3^2}\right] 
                             + p_5 \exp\left[-\frac{(t-p_7)^2}{2p_6^2}\right]
    \label{eqn:doubleGaussian}
\end{equation}

\noindent with free parameters $p_{1...7}$. For single laser pulses, 
$p_5 = 0$, and $p_6$ and $p_7$ were ignored. The true signal waveform is not
exactly Gaussian due to 
the effects of the amplifier circuitry. However, a more precise model would
be more computationally expensive, and we expect that 
the Gaussian fits will be accurate enough for the purposes of 
real-time analysis.

The parameters $p_i$ are determined in a Levenberg-Marquardt 
least-squares fitting procedure. The integrals of the fitted curves 
follow as $\sqrt{2\pi}p_2p_3$ for single pulses and
$\sqrt{2\pi}(p_2p_3+p_5p_6)$ for double pulses. The choice of whether to
fit to a single or double pulse can be made based 
on laser energy signals: if two lasers fire, a double pulse is fitted; 
otherwise, a single pulse is fitted.

The curve-fitting approach requires substantially more complex computations
than the previous approaches. Each least-squares iteration requires
at least two evaluations of the exponential function for each time point
within the integration window, as well as the inversion of an
an $n\times{n}$ matrix, where $n$ is 
the number of free parameters. 
We employed the Armadillo C++ linear algebra
library\cite{sanderson2016a,sanderson2018a} to 
ensure efficient matrix inversion.

\begin{figure}
    \begin{center}
    \includegraphics[width=0.5\textwidth]{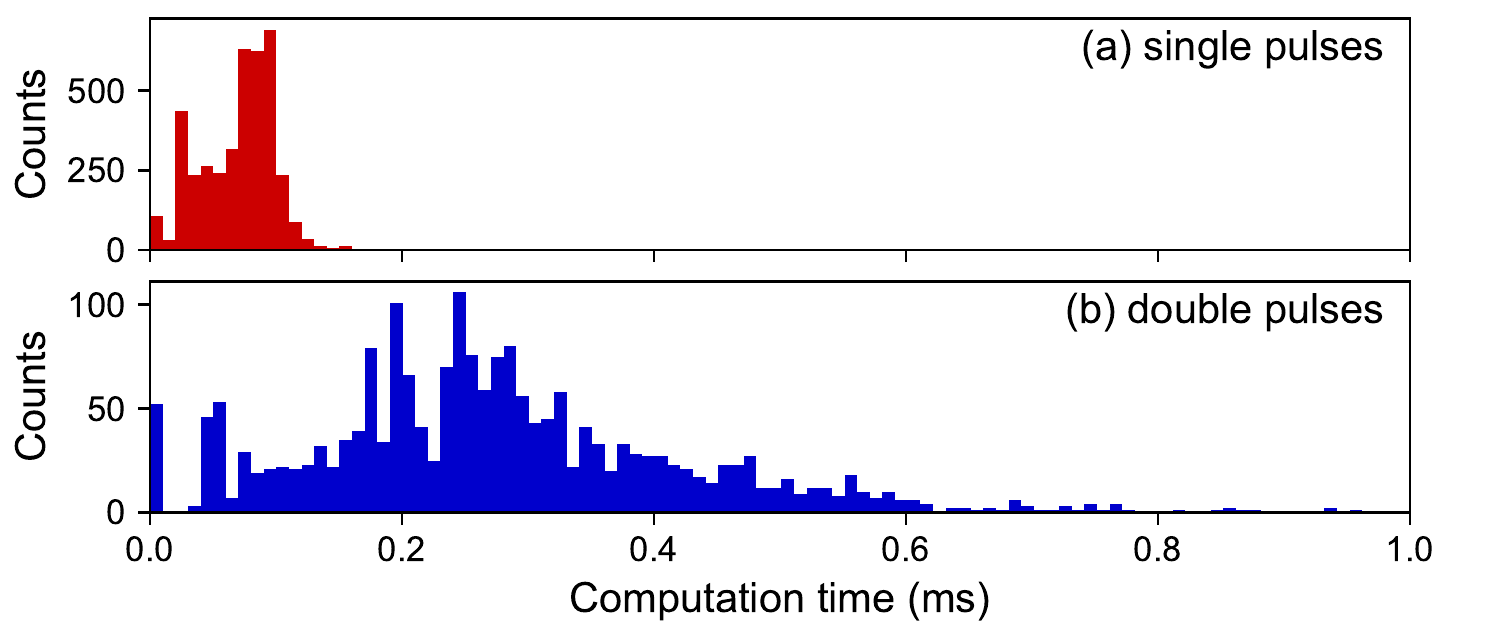}
    \caption{Histograms of the total computation time required to evaluate 
             signal amplitudes from each of five polychromator channels using
             curve fitting.
             (a) Single laser pulses; 
             (b) double laser pulses. }
    \label{fig:time_hist}
    \end{center}
\end{figure}

Evaluation times for curve-fitting during a survey of five example shots
are shown in Fig.~\ref{fig:time_hist}.
Each sample corresponds to the total time to evaluate 
fitting parameters for signals from each of five polychromator channels from a 
single scattering volume for a particular laser pulse during a given 
discharge. Values near zero correspond to
samples in which all the signals from a polychromator were
less than their respective SNR (occurring, e.g., at the beginnings and ends
of discharges), in which case the procedure would not attempt any fits.
Overall, the computation time to process five polychromator signals
was less than 110~$\mu$s in 95\% of cases for single laser pulses. For double
pulses, the total time was less than 550~$\mu$s in 95\% of cases, 
although a number of outliers required more than 800~$\mu$s.

As stated previously, the real-time system with its original software 
has an end-to-end processing time of 16.7~ms or less. 
Hence, in a worst-case performance scenario in 
which the system requires a full 16.7~ms to complete its operations for a laser
pulse, replacing the peak-detection procedure with curve-fitting 
could extend the required processing time by 0.5~ms to 1~ms. Under these 
conditions, the system could not accommodate laser pulses at the
60~Hz repetition rate for which it was originally designed. However, it could
still accommodate rates of up to 50~Hz, which would have cycle
deadlines of $\geq$20~ms. In particular, the incorporation of curve-fitting
would still work on LHD when a 30~Hz rate is employed.

\section{Temperature and density calculations}
\label{sec:ref_compare}

\begin{figure*}
    \begin{center}
    \includegraphics[width=\textwidth]{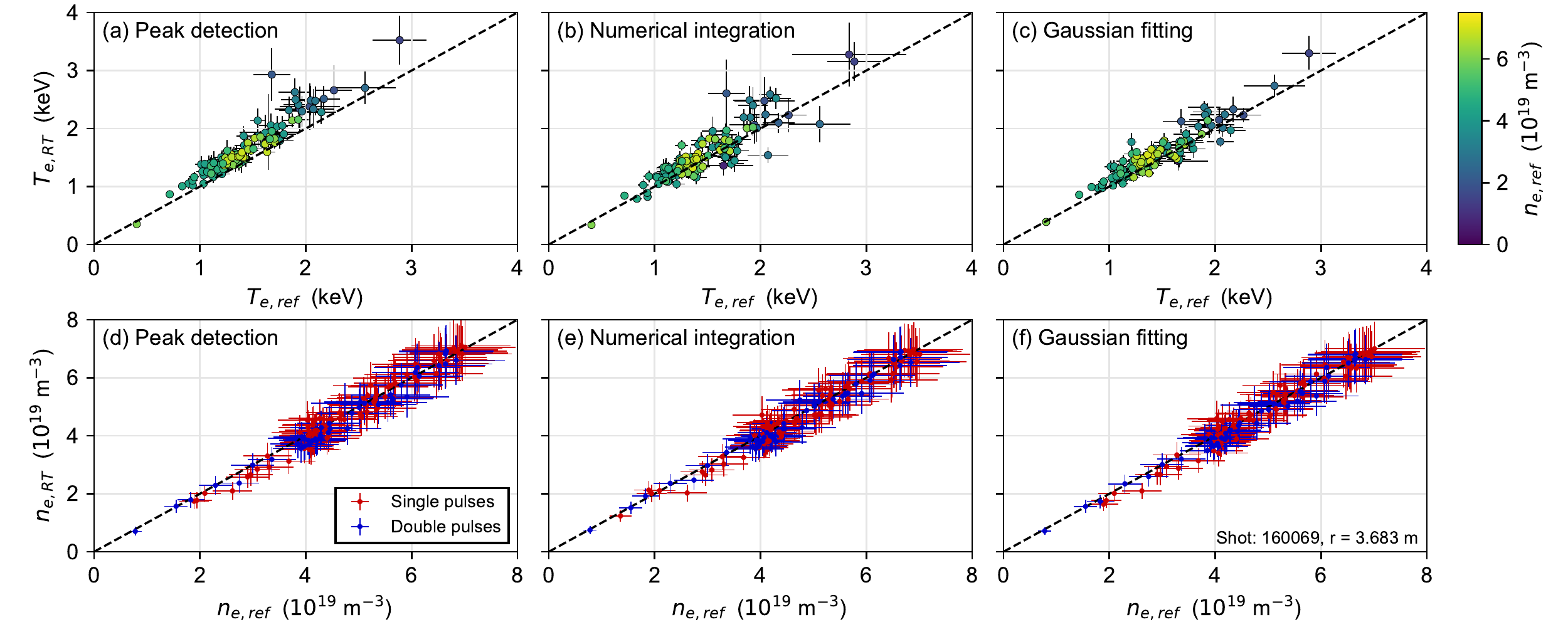}
    \caption{Comparison of $T_{e,\text{RT}}$ and $n_{e,\text{RT}}$ calculated 
             by functions from the real-time code employing different 
             signal-processing methods with reference values 
             $T_{e,\text{ref}}$ and $n_{e,\text{ref}}$ calculated by the 
             standard LHD post-processing routines.
             (a)-(c) $T_e$ with color scaling according to $n_e$;
             (d)-(f) $n_e$, colored according to pulse type.}
    \label{fig:te_ne_scatter}
    \end{center}
\end{figure*}

Fig.~\ref{fig:te_ne_scatter} compares example $T_e$ and $n_e$
estimates using the three different procedures described
in Sec.~\ref{sec:sigEval}. The reference values
$T_{e,\text{ref}}$ and $n_{e,\text{ref}}$ were obtained from the standard LHD 
post-processing routines. The real-time functions calculated $n_e$ using the 
values of $c$ (Eq.~\ref{eqn:chi2}) scaled and offset by calibration factors 
determined through fitting to $n_{e,\text{ref}}$ obtained from a separate set of
discharges, independently for single and double laser pulses. Points with 
$n_e<0.5\times10^{-19}$~m$^{-3}$ and $T_e>4$~keV generally had high
uncertainty for this example and were omitted from the plot.

Overall, the values of $T_e$ and $n_e$ as computed by all three methods track
well with the reference values, almost always agreeing to within 0.5~keV and
$0.5\times10^{19}$~m$^{-3}$. In general, the real-time software tends to 
slightly
overestimate $T_e$ relative to the reference values.
Note that the overestimation of $T_e$ is most severe for the peak-detection
method (Fig.~\ref{fig:te_ne_scatter}a), an effect that we believe is related 
to the systematic errors in the spectral fitting for this method.
The values of $T_e$ determined from numerical integration exhibit greater
scatter than the other methods due to the impacts
of noise on the integration: the standard deviation of $T_e-T_{e,\text{ref}}$
was $190$~eV for the numerical integration approach, whereas it was
$160$~eV for the pulse amplitude approach and $130$~eV for curve fitting. 
The tendency of curve-fitting methods to produce $T_e$ values with 
less scatter than direct integration has also been observed, for example,
in the Thomson scattering system at KSTAR.\cite{lee2017a}

\section{Conclusions}

In summary, we have adapted and commissioned a real-time evaluation system
for Thomson scattering on the LHD experiment, representing the first
\textit{in situ} operation of the system.
We have implemented and compared three different approaches to evaluating the
polychromator signals, each of which offers advantages and disadvantages. 
The curve fitting approach generally provides the greatest accuracy and
precision, but at the expense of extra computation time.
Nevertheless, initial comparisons of $T_e$ and
$n_e$ output by all three methods mostly exhibit good agreement with values 
calcuated by the standard post-processing software. 

\begin{acknowledgments}
The authors would like to thank B.~P.~LeBlanc for the helpful discussions.
This work was supported by the US Department of Energy under contracts
DE-AC02-09CH11466 and DE-SC0015480, as well as the NIFS LHD Project under 
Grant No.~ULHH040. The work was conducted within the framework of the 
NIFS/PPPL International Collaboration. The US Government retains a 
non-exclusive, paid-up, irrevocable, world-wide license to publish or reproduce 
the published form of this manuscript, or allow others to do so, for 
US Government purposes.
\end{acknowledgments}

\section*{Data availability}
The data supporting the findings of this study are available at 
\hyperref[http://arks.princeton.edu/ark:/88435/dsp014t64gr25v]{http://arks.princeton.edu/ark:/88435/dsp014t64gr25v}.

%

\end{document}